\documentclass{article}
\usepackage{amsmath}
\usepackage{booktabs}
\usepackage{multirow}
\usepackage[T1]{fontenc}
\usepackage[utf8]{inputenc}
\usepackage[]{ismir} 
\usepackage{amsmath,cite,url}
\usepackage{graphicx}
\usepackage{color}
\usepackage{bbm}

\title{Beyond Genre: Diagnosing Bias in Music Embeddings Using Concept Activation Vectors}

\threeauthors
  {Roman B. Gebhardt} {Cyanite \\ \texttt{roman@cyanite.ai}}
  {Arne Kuhle} {Cyanite \\ \texttt{arne@cyanite.ai}}
  {Eylül Bektur} {Cyanite, TU Berlin \\ \texttt{bektur@campus.tu-berlin.de}}

\def\authorname{Roman B. Gebhardt, Arne Kuhle, and Eylül Bektur}

\begin{document}

\maketitle

\begin{abstract}
Music representation models are widely used for tasks such as tagging, retrieval, and music understanding. Yet, their potential to encode cultural bias remains underexplored. In this paper, we apply Concept Activation Vectors (CAVs) to investigate whether non-musical singer attributes - such as gender and language - influence genre representations in unintended ways. We analyze four state-of-the-art models (\textit{MERT}, \textit{Whisper}, \textit{MuQ}, \textit{MuQ-MuLan}) using the STraDa dataset, carefully balancing training sets to control for genre confounds. Our results reveal significant model-specific biases, aligning with disparities reported in MIR and music sociology. Furthermore, we propose a post-hoc debiasing strategy using concept vector manipulation, demonstrating its effectiveness in mitigating these biases. These findings highlight the need for bias-aware model design and show that conceptualized interpretability methods offer practical tools for diagnosing and mitigating representational bias in MIR.
\end{abstract}

\section{Introduction}\label{sec:introduction}

Model bias is a well-known challenge across machine learning domains. While extensively studied in NLP and computer vision~\cite{Garrido-Munoz:21, Ntoutsi:20}, it has recently also gained growing attention in MIR~\cite{Holzapfel:18, Shakespeare:20, Wang:23}. Beyond degrading performance, model bias also poses serious challenges for ML fairness~\cite{Holzapfel:18}. In MIR, models may learn to reflect or amplify societal imbalances present in the music industry, reinforcing stereotypes in classification, recommendation, and musical understanding. This phenomenon has been empirically demonstrated in prior work on artist gender bias in music recommendation systems~\cite{Shakespeare:20}.
Recent research in explainable AI for MIR has shown that multi-modal large language models (LLMs) trained for music understanding often struggle to utilize audio content, at times ignoring it entirely in favor of accompanying text data \cite{Weck:24}. This overreliance on the textual modality can lead to auditory hallucinations, where models generate plausible-sounding but inaccurate descriptions. We hypothesize that such failures may also partially stem from biased internal audio representations learned by neural models, where stereotypical musical patterns or demographic correlations overly influence the internal encoding. Already flawed audio representations will naturally hinder the model’s ability to generalize and reason faithfully about music. We therefore shift our focus to analyzing the audio representations themselves, where such biases may originate but remain largely unexplored. To address this gap, we employ Concept Activation Vectors (CAVs) \cite{Kim:18} to systematically probe for unwanted concept entanglement.
We investigate whether non-musical factors like singer gender and language influence genre representations. While these attributes should not affect genre representation, we hypothesize that skews in training data lead models to associate them with specific genres. Prior work suggests such imbalances are genre-dependent~\cite{Richardson:22, Tabak:23, EppsDarling:20, Ferraro:20}: \textit{Metal} and \textit{Hip-Hop} are heavily male-dominated, while genres like \textit{Pop}, \textit{Electronic}, and \textit{R\&B} are more balanced~\cite{EppsDarling:20}. A model might thus associate male vocals with \textit{Metal} and female vocals with \textit{Pop}, even though vocal gender is not a genre-defining trait. Similarly, while language may serve as a genre cue in specific cases (e.g., Portuguese in \textit{Brazilian music}), overreliance on dominant languages risks marginalizing others~\cite{Howard:11}.
To investigate these biases, we quantify how strongly non-musical attributes are reflected in the model’s latent space. By adapting Testing with CAVs (TCAV) \cite{Kim:18} for frozen audio encoders, we estimate how consistently genre-specific audio embeddings align with a given concept direction. Secondly, we explore the possibilities of applying CAVs for concept removal or addition, representing a simple post-hoc de-biasing strategy. We publish our code on Github.\footnote{\url{https://github.com/WhoCares96/CAV-MIR}}

This work aims to provide insights into how state-of-the-art music representation models encode and propagate bias, advocating for more fairness-aware design in MIR. While we focus on audio encoders, our approach generalizes to other music-related models. It offers a lightweight, interpretable framework to surface and mitigate biases using small, targeted datasets.

\section{Related Work}\label{sec:related_work}

\subsection{Concept-based Explanations}

Concept-based explanations have emerged in machine learning as a way to make models more interpretable by aligning their latent representations with human reasoning and intuitive concepts, rather than solely relying on low-level input features such as individual pixels or raw data points \cite{Yeh:22}. Concept-based interpretability for neural networks typically follows two main approaches: (1) designing neural network architectures that are intrinsically interpretable and (2) generating post-hoc explanations for already-trained networks.  Intrinsic, concept-aware methods from the first category typically achieve high task accuracies, link concepts directly to predictions through learning which enable more effective interventions of concepts on the model’s decisions \cite{Zhang:24, Koh:20}. However, these intrinsic approaches require training with labelled concept data, which can be costly to obtain \cite{Koh:20}. 

In contrast, post-hoc concept-based methods, notably the Concept Activation Vector (CAV) method introduced by Kim et al. \cite{Kim:18}, have proven particularly effective due to their explicit alignment with human-understandable and domain-relevant concepts without the need of annotated concept labels or model retraining \cite{Foscarin:22, Thakoor:21, Erogullari:25}. Recent research applies CAV-based methods to verify whether models focus on desired semantic concepts, like object shapes, or undesired spurious signals \cite{Anders:22}. Therefore, CAV-based methods have had successful applications in explainability, understanding model representations, concept entanglement detection, spatial dependency evaluations \cite{Nicolson:25} and bias evaluation \cite{Wei:21}. The related Testing with CAV (TCAV) method quantifies the influence of each concept on the model’s predictions by computing directional derivatives along the corresponding CAVs \cite{Kim:18}.

In the MIR domain, researchers have applied post-hoc interpretability methods that operate directly on the input spectrogram, perturbing small time–frequency patches and tracking the resulting output changes to reveal which regions most strongly influence the classifier’s decision \cite{Mishra:17}. These methods typically lack the broader conceptual insights that methods such as CAVs offer by directly associating predictions with human-defined concepts \cite{Kim:18} and providing explanations of model behaviour provided by TCAV. 

Motivated by this gap, recent MIR work incorporates concept-based methods, such as CAVs, for structuring complex genre and mood categories into hierarchical representations \cite{Afchar:22}, and generating explanations tailored explicitly to musicologists by visualizing interpretable musical concepts \cite{Foscarin:22}. Building upon these developments, our work applies CAV-based approaches that explicitly align model bias with human-defined concepts. This follows methodologies introduced for bias exploration and for debiasing retrieval systems with CAVs.

\subsection{Bias Exploration and Mitigation}
Existing studies interpret bias via individual neurons, higher-dimensional subspaces, or linear directions in latent space \cite{Yu:25, Krishnan:24, Bolukbasi:16}. By adapting CAV, we employ the linear-directions approach, defining concepts as linear directions learned through supervised training on model activations. 

Recent research has utilized CAV methods particularly in computer vision to detect and quantify undesirable model reliance on demographic biases \cite{Correa:24, Tong:20}. To the best of our knowledge, the speech-audio TCAV/CAV literature has thus far focused mainly on accent-related bias, with English as the input language \cite{Wei:21}.  In contrast, our research explicitly examines biases arising from differences across multiple spoken languages in audio data, making it the first to explore language-related bias across both speech and MIR domains.

Methods concerned with bias in the audio domain can be grouped into techniques applied before or during model training, and post-hoc approaches aimed at identifying and mitigating biases after training, primarily through debiasing latent representations. 
Pre-or-during strategies include mitigation of data and annotation biases including careful dataset selection and improved transparency through detailed documentation \cite{Maia:24}, counterfactual attention learning \cite{Lin:25}, and token masking to prevent overfitting to the dataset during learning \cite{Zhao:25}. 
Post-hoc methods in MIR that employ bias exploration utilize statistical significance tests to compare performance distributions across affected and advantaged groups \cite{Yesiler:22}, and analyze performance disparities by evaluating differences between universal and culturally adapted models \cite{Holzapfel:14}.

Our approach in this paper is most similar to that of \cite{Wang:23}, who apply a dimensionality-reduction method, Linear Discriminant Analysis (LDA), to identify the bias direction in pre-trained audio embeddings by training LDA to separate two different datasets. Wang et al. \cite{Wang:23} address the dataset bias that is influenced by both the training approach of the embeddings and the alignment of class vocabularies between datasets. In contrast, our CAV-based method defines undesirable biases as concepts which allows us to analyze high-level biases related to how the artist representations manifest themselves in the embeddings. Whereas Wang et al. \cite{Wang:23} remove domain‐sensitivity bias by subtracting a linear projection from the embedding itself, our approach instead adjusts the linear CAV classifier that probes the embedding. Both strategies are linear, but they act at different stages of the pipeline. 

Therefore, our work expands upon existing post-hoc bias exploration and retrieval-system debiasing efforts by applying CAV-based interpretability to address and mitigate the effects of demographic and sociocultural factors in music representation models.

\section{Music Representation Models}\label{sec:models}

We evaluate four state-of-the-art music representation models in their publicly available, pretrained form: \textbf{MERT}, \textbf{Whisper}, \textbf{MuQ}, and \textbf{MuQ-MuLan}. All four can be used to generate audio embeddings for tasks such as \textit{music tagging}, \textit{zero-shot and downstream classification}, \textit{music retrieval}, and as \textit{audio encoders} in music understanding LLMs.

\textbf{MERT} (MERT-v1-95M) \cite{Li:23} is a self-supervised Transformer model trained on large-scale music datasets using masked modeling and pseudo-labels from acoustic and musical teacher models. It captures musical structure and semantics, making it particularly effective for tasks like music-text retrieval and genre classification. It is widely used in open-source Music-LLMs \cite{Weck:24}.

\textbf{Whisper} (Whisper-large-v2) \cite{Radford:22} is a speech recognition model trained on 680,000 hours of multilingual and multitask audio data. While originally designed for \textit{automatic speech recognition (ASR)}, it has shown some capability in processing music audio, particularly for transcription tasks \cite{Zhuo:23}. Like MERT, Whisper is often used as an encoder in music LLMs pipelines \cite{Weck:24}, contributing to music retrieval and captioning tasks. Unlike the other models, Whisper was not trained with music; instead, it is optimized to capture linguistic structure, phonetic features, and prosody, which presumably could lead to less entanglement between language and musical genre as the model's main focus should be the singers' voice instead of the musical content.

\textbf{MuQ} (MuQ-large-msd-iter) \cite{Zhu:25} is a self-supervised model that learns discrete music representations via Mel Residual Vector Quantization (Mel-RVQ). It is trained \textit{solely on audio data}, without manual labels, and excels at zero-shot tagging and instrument classification. 

\textbf{MuQ-MuLan} (MuQ-MuLan-large) \cite{Zhu:25} extends MuQ by incorporating a \textit{joint training objective with a text encoder (MuLan)} on a large-scale corpus of paired music--text data. This allows MuQ-MuLan to align musical and textual features in a shared embedding space, enabling music--text retrieval and captioning.

In our study, we include both MuQ and MuQ-MuLan to assess how text supervision affects concept encoding and entanglement. While MuQ captures structure-driven representations grounded in audio alone, MuQ-MuLan - through its alignment with text - may encode stronger correlations between musical and non-musical attributes, potentially leading to increased cultural or linguistic bias.

\section{Dataset}\label{sec:dataset}

We use STraDa (Singer Traits Dataset) \cite{Kong:2024}, a large-scale dataset designed for analyzing singer-related attributes in music. Specifically, we leverage the automatic-strada subset, which includes metadata for over 25,000 tracks. This metadata - covering lead singer gender (male, female, non-binary) and language is cross-validated across multiple sources to ensure reliability.
To obtain audio, we use the Deezer API to retrieve 30-second audio previews and successfully collect 22,168 tracks. To address underrepresentation of certain genre–gender combinations in STraDa, we supplement the dataset with 251 additional tracks from Deezer playlists specifically curated around the underrepresented concepts.\footnote{Affected genres are marked with * in Figure \ref{fig:tcav-combined}} These playlists provide an external, non-manual source of curation. For quality assurance, based on playlist name and listening we manually annotate each track's genre, and the singer's assumed gender and language, discarding mismatches with the intended playlist theme. Acknowledging potential curation bias, we publicly release all playlist IDs, track IDs, and metadata in the additional material.

From this corpus, we construct balanced train / test datasets for nine binary classification tasks: gender (male, female) and the seven most common languages in STraDa (en, fr, it, pt, ja, es, de). Training sets are used to learn CAVs; test sets to compute TCAV scores.  Following \cite{Kim:18}, we consider poorly projecting CAVs as incapable of reliably representing a concept and thus unsuitable for bias analysis.

To construct the CAV training and test sets, we stratify the data across all combinations of language, genre, and gender. For each binary concept (e.g., gender or language), we sample an equal number of positive (e.g., gender = female; language = Portuguese) and randomly selected non-positive samples (e.g., gender $\neq$ female; language $\neq$ Portuguese) within each genre to prevent genre-based confounds. When data is abundant, we fix a maximum of 50 samples per (label, genre) cell for training and assign the remainder to testing; when scarce, we downscale proportionally. This yields genre-balanced, disjoint train/test splits with broad subgroup diversity. To reduce concept entanglement, we cap the number of training samples per (concept, genre) pair, preventing dominant subgroups from overwhelming the concept definition.

Such careful balancing is essential to ensure that observed bias reflects the model’s internal representation - not artifacts of the dataset. While our strategy controls for known attributes using available metadata, it cannot eliminate latent or unobserved factors. For example, female- and male-led English-language jazz may differ in timbre or arrangement, but we assume they remain acoustically comparable in terms of genre-defining traits. Our CAVs aim to isolate the intended concept under this approximation, acknowledging potential residual entanglement.

\section{Method}\label{sec:method}

CAVs assume that a target concept is \textit{linearly separable} in a model’s latent space - that is, there exists a hyperplane that distinguishes between samples with and without the concept. We construct a CAV by training a logistic regression model on the final-layer latent embeddings $\mathbf{x}$, which learns the decision function:

\begin{equation} 
\hat{y} = \sigma(\mathbf{w}^\top \cdot \mathbf{x} + b)
\end{equation}
where $\sigma$ is the sigmoid function. The hyperplane $\mathbf{w}^\top \cdot \mathbf{x} + b = 0$ defines the decision boundary, and the CAV is the normal vector $\mathbf{w}$ orthogonal to this boundary. To validate its reliability, we can evaluate the classifier’s accuracy. High performance indicates that the concept is linearly encoded; otherwise, the CAV is considered unreliable. This does not necessarily mean the concept is absent from the embedding space, but rather that it cannot be fully captured by our linear analysis. Once a CAV is learned, it can be used to measure how strongly individual audio samples align with a given concept direction. This enables intuitive inspection of model behavior  -  for example, identifying whether certain genres systematically reflect demographic attributes. To quantify such patterns, we adapt the TCAV approach to estimate how often a category’s embeddings align with the concept, which may indicate potential bias in the representation.

\subsection{Measuring Concept Alignment Using TCAV}

TCAV \cite{Kim:18} provides a statistical framework to quantify the extent to which a model’s internal representations rely on a given concept, enabling structured and scalable analysis of concept influence. 

Unlike the original TCAV method - which computes directional derivatives of class logits from a trained classifier - our approach slightly deviates, as it operates on frozen audio encoders without access to gradients from a softmax layer or any downstream prediction head. 
Instead, we evaluate alignment with a concept by testing whether genre-specific final-layer embeddings fall on the positive side of the learned hyperplane.  We omit the sigmoid and use its logit as the projection,
\begin{equation} 
p_{\text{CAV}}(\mathbf{x}) = \mathbf{CAV}^\top \cdot \mathbf{x} + b
\label{eq:p_cav}
\end{equation}
This projection reflects the sample's alignment with the learned concept. Including the bias term $b$ is essential here - unlike in the original TCAV formulation - since we do not compute directional derivatives, where the bias would cancel out. Instead, we interpret the CAV as a complete decision function. 

Our TCAV score is then defined as the fraction of samples with a positive projection, indicating how often the model’s representations align with the concept:
\begin{equation} 
\text{TCAV} = \frac{1}{N} \sum_{i=1}^{N} \mathbbm{1}(p_{\text{CAV}}(\mathbf{x}_i) > 0)
\end{equation}
where $\mathbbm{1}$ denotes the indicator function, returning 1 if the condition is true and 0 otherwise. By comparing TCAV scores across different genres, we assess the extent to which the model’s genre representations relate to the given concept. Because the concept test sets are balanced, the expected TCAV score under the null hypothesis (no alignment) is $0.5$. Scores significantly above or below this threshold indicate systematic alignment - and thus potential bias.
To ensure the robustness of our findings, we follow the original TCAV protocol: for each concept, we train 500 CAVs on independently sampled, balanced training subsets comprising 25\% of the data. This yields a distribution of TCAV scores per concept–genre pair. We then conduct a two-sided t-test to evaluate whether the mean TCAV score significantly deviates from $0.5$, and apply a Bonferroni correction to account for multiple comparisons.

\subsection{Adjusting Bias via Concept Vector Manipulation}

To explore and mitigate bias in genre representations, we modify genre-specific CAVs using bias-related concept directions. As a case study, we focus on \textit{Hip-Hop}, which we expect to be negatively aligned with the \textit{Female vocal} concept.  We create a Hip-Hop CAV based on a gender-balanced dataset and rank tracks in a similarly balanced test set using their \textit{Hip-Hop} projection scores $p_{\text{CAV}}(\mathbf{x})$ (Eq.~\ref{eq:p_cav}). If bias is encoded, male-vocal tracks should appear near the top.
To attenuate this effect, we adjust the \textit{Hip-Hop} CAV by interpolating toward the \textit{Female vocal} concept:
\[
\mathbf{CAV}_{\text{hiphop}}^{\text{adj}} = (1 - \lambda) \cdot \mathbf{CAV}_{\text{hiphop}} + \lambda \cdot \mathbf{CAV}_{\text{female}},
\]
with $\lambda \in [0, 1]$ controlling the adjustment strength. This shifts the decision boundary toward the female vocal direction, reducing male dominance in the top-ranked tracks. Subtracting the \textit{Male vocal} CAV should achieve a similar effect, given that the two concept vectors are approximately opposites by construction.

This procedure reveals how semantically meaningful directions can influence genre alignment and provides a simple, reproducible post-hoc technique to analyze and mitigate representational bias. While we adjust CAV directions here, similar strategies could in principle be applied to embeddings directly.

\section{Results}\label{sec:results}

\begin{figure*}[t]
    \centering
        \centering
        \includegraphics[width=\textwidth]{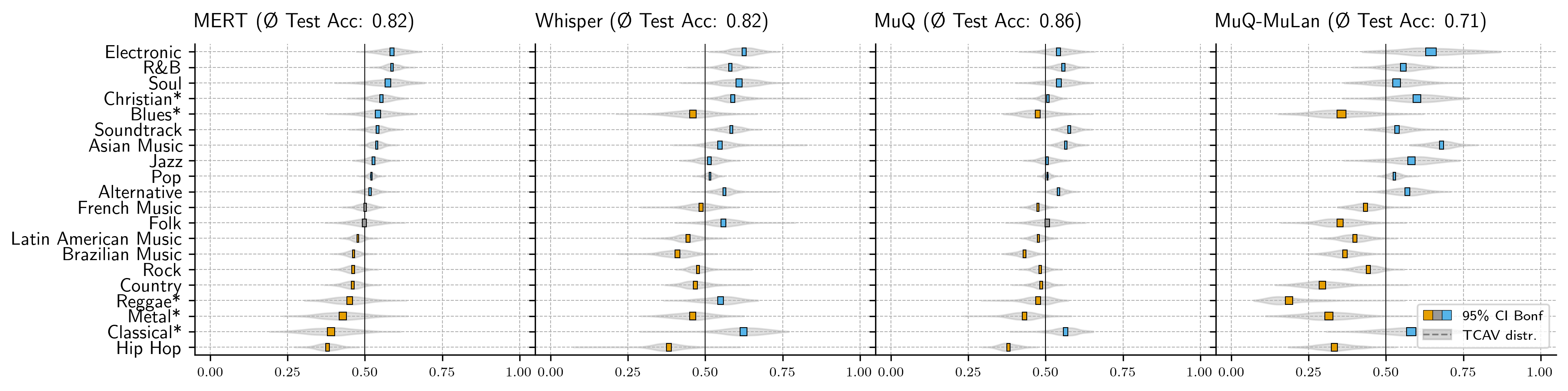}
        \label{fig:tcav-female}
    
    \vspace{0em}
    
        \centering
        \includegraphics[width=\textwidth]{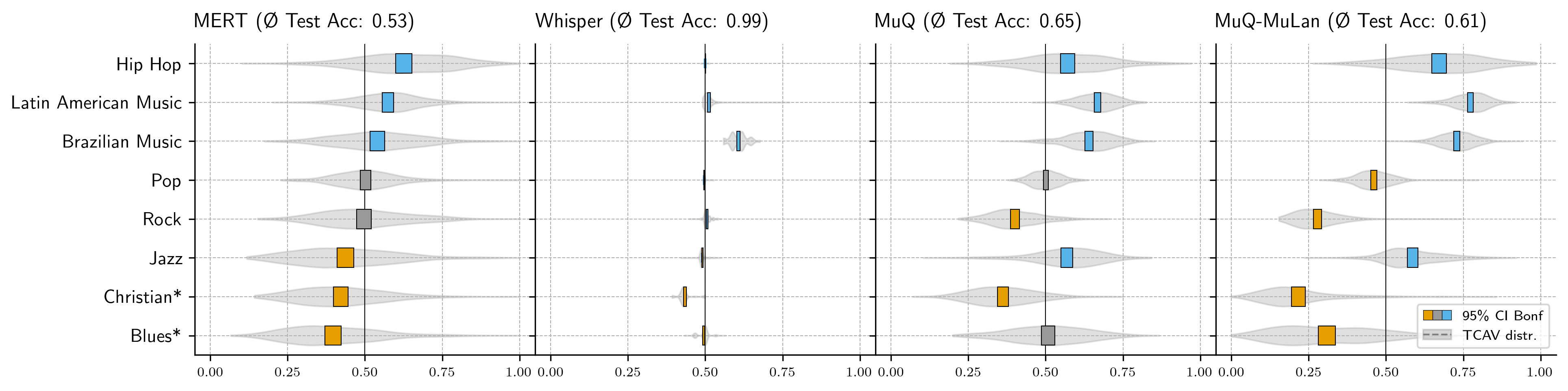}
        \label{fig:tcav-portuguese}
    
    \caption{Per-model TCAV evaluation for concepts \textbf{Gender-Female} (upper) and  \textbf{Language-Portuguese} (lower). Violin plots represent the distribution of TCAV scores across 500 CAV trainings. Boxes indicate a Bonferroni-corrected 95\% confidence interval. Blue/orange coloring indicates significant positive/negative correlation between genre and concept.}

    \label{fig:tcav-combined}
\end{figure*}

To investigate the influence of non-musical attributes on genre representation, we analyze the model responses to two representative concepts in depth: \textbf{Female vocals} and \textbf{Portuguese language}. The two discussed concepts were selected for their illustrative power - \textit{Female vocals} as a proxy for the singer's gender (noting that the \textit{Male vocals} concept yields largely inverse results), and \textit{Portuguese} as a representative example of linguistic variation in music. \textit{Portuguese} was specifically chosen due to its strong association with genres such as \textit{Latin American Music} and \textit{Brazilian Music}, where meaningful entanglement might be expected, in contrast to other genres where such associations should be less likely. Additional results for all concepts are provided in the supplementary material and largely mirror the trends described in the following sections.

\subsection{Evaluation of the Female Concept}

We first assess the TCAV scores across genres for the concept of \textit{Female vocals} across the four models, displayed in Figure \ref{fig:tcav-combined}. The average classification accuracy of the trained CAVs exceeds 80\% for all models except MuQ-MuLan, indicating that the concept is linearly encoded in their latent spaces and thus reliably captured by the CAVs.
Most TCAV scores deviate significantly from the chance level, indicated by the Bonferroni-corrected 95\% confidence intervals not spanning across the 0.5 mark. This suggests the presence of bias in the internal genre representations of all models. The fact that these scores often diverge in direction between models - despite being trained on the same balanced data - strongly suggests that the observed biases reflect genuine differences in how each model encodes the concept.

\textbf{MERT} displays significant negative biases for genres such as \textit{metal}, \textit{rock}, and \textit{Hip-Hop}, with TCAV scores substantially below 0.5. In contrast, genres like \textit{Electronic}, \textit{R\&B}, and \textit{Soul} show clear positive associations with the \textit{Female vocals} concept. These patterns align with common vocal stereotypes associated with these genres. Interestingly, \textit{Classical} music reveals the second-strongest negative bias in MERT, despite showing a significantly positive association with female vocals in all other models.

\textbf{Whisper} demonstrates a notably different pattern. While it agrees with expectations for a few genres (e.g., \textit{Metal}, \textit{Hip-Hop}), its TCAV scores diverge - sometimes substantially - in other genres. These inconsistencies suggest that Whisper’s internal representations differ from those of MERT. One plausible explanation is Whisper’s origin as an automatic speech recognition (ASR) model, which may render it more sensitive to vocal characteristics such as pitch, timbre, or even lyrics, leading to model-specific associations between vocal traits and genre.

\textbf{MuQ}'s distribution patterns interestingly resemble those of Whisper, with both models showing relatively subtle genre-specific deviations. Neither model is exposed to music-specific textual labels during training. This absence of genre- or culturally-aligned text input may encourage more structurally grounded representations, resulting in similar concept entanglement with non-musical attributes.

\textbf{MuQ-MuLan}, while aligned in general bias direction with its sibling MuQ, reveals much stronger and more polarized TCAV scores. The model shows significantly negative scores for many genres, and amplified positive scores in others. This suggests that MuQ-MuLan, trained with text supervision, encodes stronger cultural associations between gender and genre. Notably, MuQ-MuLan exhibits strong and statistically significant TCAV scores despite lower CAV test accuracy. This suggests that the model’s representations are aligned with the concept in a more entangled or diffuse manner - capturing meaningful bias even when the concept is not cleanly linearly separable.
In summary, all models exhibit genre-specific biases related to female vocals, with significant variation in both magnitude and direction. These findings suggest that the training objective, modality, and supervision signal have a substantial influence on how gendered information becomes encoded, and underscore the need for awareness of such effects in downstream MIR applications.

\subsection{Evaluation of the Portuguese Language Concept}

We now turn to the TCAV evaluation of the \textit{Portuguese language} concept across genres (Figure~\ref{fig:tcav-combined}). While Whisper's accuracy is close to perfect, the average CAV classification accuracy of the other models is lower than for the gender-based concepts, and while still above chance level it should be interpreted with caution.

\textbf{MERT} shows strong positive TCAV scores for \textit{Latin American Music} and \textit{Brazilian Music}, which aligns with expectations given the natural linguistic-cultural overlap. Surprisingly, MERT shows the strongest bias for \textit{Hip-Hop}. In contrast, genres such as \textit{Rock}, \textit{Christian}, and \textit{Blues} exhibit significant negative biases, suggesting a possible entanglement of Portuguese with stylistic or rhythmic features more prevalent in other genres.

\textbf{Whisper}, by contrast, yields TCAV scores that remain close to the null hypothesis of 0.5 for most genres, indicating relatively little concept alignment. This supports the interpretation that Whisper - trained primarily for multilingual speech recognition - encodes language in a more disentangled fashion, possibly ignoring most musical information. A notable exception is a small but significant positive bias for \textit{Brazilian Music}, hypothetically reflecting Whisper's heightened sensitivity to vocal or phonetic traits present in Brazilian Portuguese singing.

\textbf{MuQ} captures the expected positive associations between Portuguese and the \textit{Latin American} and \textit{Brazilian Music} genres, while showing negative or neutral biases for other genres. These biases are more strongly amplified in \textbf{MuQ-MuLan}, where all genre-specific TCAV scores exhibit clearer polarity. This again may highlight the effect of multimodal training: while MuQ learns from acoustic features alone, MuQ-MuLan's text alignment appears to reinforce cultural and linguistic correlations, magnifying concept entanglement.

\subsection{Concept Debiasing}

\begin{figure}[t]
  \centering
  \includegraphics[width=0.9\linewidth]{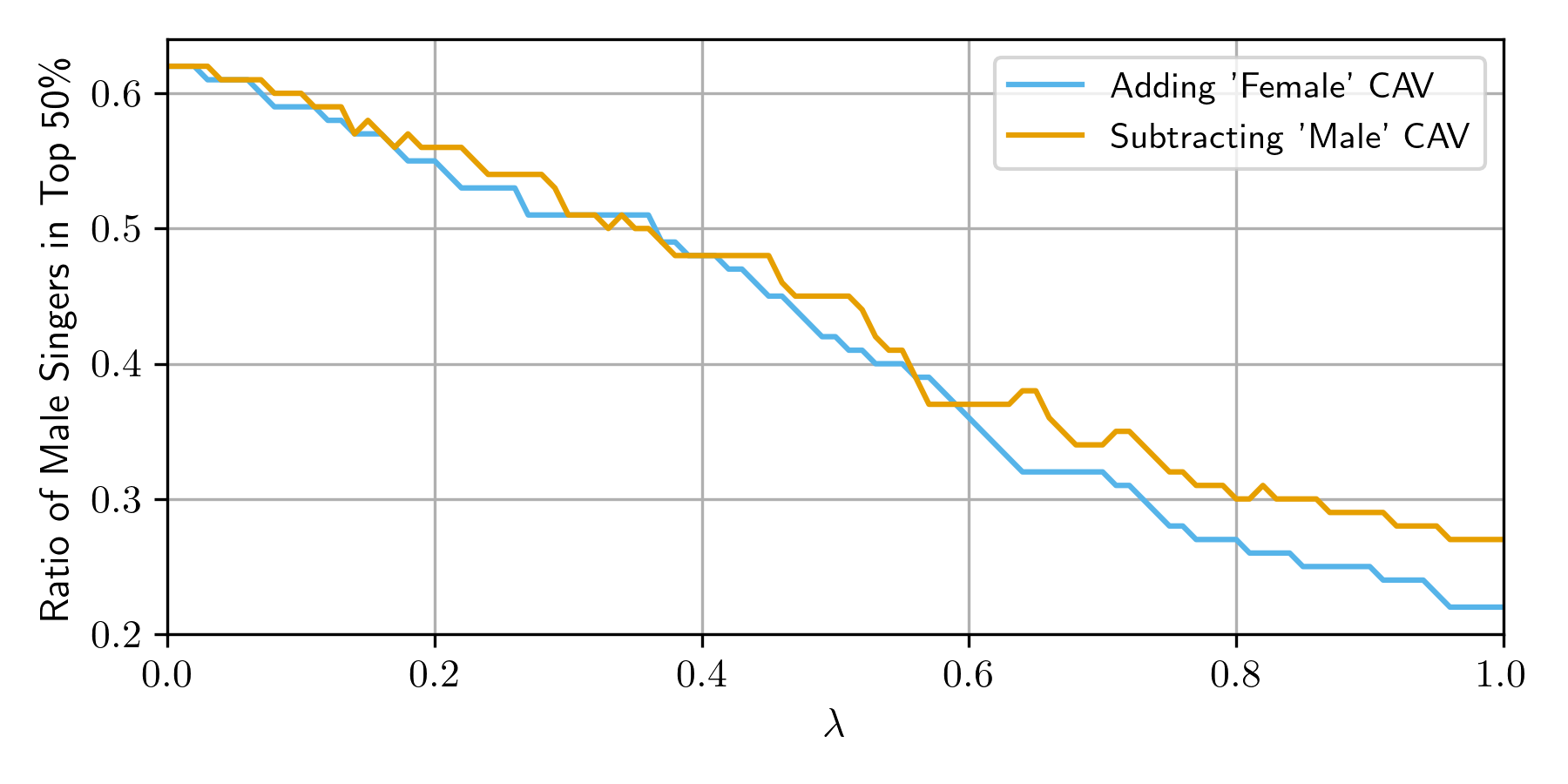}
  \caption{Effect of concept vector debiasing on the \textit{Hip-Hop} CAV. We show the ratio of male singers among the top 50\% of ranked tracks when gradually adding the \textit{Female vocals} CAV (blue) or subtracting the \textit{Male vocals} CAV (orange), as a function of the adjustment weight $\lambda$.}
  \label{fig:hip_hop_debiasing}
\end{figure}

Figure \ref{fig:hip_hop_debiasing} visualizes the effect of our vector-based debiasing strategy applied to the \textit{Hip-Hop} CAV for \textbf{MuQ-MuLan}. Here, $\lambda=0$ represents the original CAV-based sorting, where a strong male overrepresentation is observed in the top 50\%, as expected from the earlier TCAV analysis. As $\lambda$ increases, we either add the \textit{Female vocals} CAV or subtract the \textit{Male vocals} CAV, and monitor the proportion of male singers among the top 50\% ranked tracks in a gender-balanced Hip-Hop test set.

Notably, both operations lead to a nearly linear reduction in male dominance as $\lambda$ increases, suggesting a meaningful semantic direction in the latent space. This indicates that the \textit{Male} and \textit{Female vocals} CAVs encode well-isolated representations of vocal gender. The consistent effects across both directions support their intuitive symmetry - expected due to their mutually exclusive and balanced construction in training. A qualitative review of male-labeled tracks that remained highly ranked after debiasing revealed that many were in fact wrongly labeled, and instead feature female vocals, further reinforcing the reliability of the learned CAVs in capturing vocal gender.
Our findings highlight that the learned CAVs capture meaningful and robust concept directions, and that concept vector manipulation can serve as a simple post-hoc strategy for adjusting model behavior, as showcased here in a ranking scenario.

\section{Limitations and Future Work}\label{sec:limitations}

Concept disentanglement is essential for interpretable representations. In our work, we mitigate spurious associations between non-musical concepts (e.g., gender, language) and genre by carefully balancing datasets to avoid subgroup overrepresentation. However, even with this balancing, learned CAVs may still be entangled with latent or unobserved factors, especially when the target concept is not cleanly separable in the embedding space. Noisy or ambiguous concept labels may further degrade the clarity of the resulting CAVs. As a result, TCAV scores may reflect not only the intended concept but also correlated dimensions, limiting interpretability. Future work could explore orthogonal CAV training strategies (e.g., \cite{Erogullari:25}) to better isolate individual concepts by explicitly reducing overlap between concept vectors in the latent space. Additionally, our linear analysis assumes a roughly interpretable embedding geometry, which may not capture complex concept interactions. Simple, acoustically salient concepts (like gendered voice timbre) may yield clearer, more interpretable CAVs than abstract, culturally embedded ones (like the singers' age or regional associations). This could unintentionally bias the analysis toward more acoustically grounded attributes.

\section{Conclusion}\label{sec:conclusion}

We systematically investigated non-musical bias in state-of-the-art music embedding models using Concept Activation Vectors (CAVs) and an adapted TCAV pipeline. Our results reveal significant and meaningful entanglements between genre representations and attributes such as singer gender and language, with variation across models. These patterns reflect known disparities in the music industry and highlight the need for bias-aware model development in MIR. Beyond diagnostic insights, we demonstrate that CAVs can serve as an intuitive and lightweight tool for post-hoc debiasing through concept vector manipulation. Our approach generalizes beyond music representation models: it can be readily applied to any MIR system that produces latent embeddings, including genre classifiers, taggers, or retrieval models. Crucially, it requires only a small set of curated concept examples, making it practical and accessible for real-world deployment. 

With this work, we aim to encourage broader research into how MIR models understand and represent music - and the social and cultural implications that follow. While we use concept-based analysis, regardless of method, our primary goal is to foster critical reflection on the biases and assumptions embedded in music technologies.

\section{Ethics Statement}\label{sec:ethics}

This work investigates representational biases in music embedding models, focusing on demographic and linguistic attributes. Our goal is to expose how models may encode and propagate social and cultural imbalances, aiming to promote fairer and more inclusive MIR systems.

We acknowledge that concepts like gender and language are complex, fluid, and socially constructed. Our binary treatment of gender (male/female) reflects limitations in available metadata and is not an endorsement of reductive framings. We recognize the broader spectrum of gender identities and emphasize the need for more inclusive data collection practices in future research. The absence of non-binary classes in our study is due to insufficient annotated data, and we encourage the community to expand upon these axes with more representative datasets.
In our data augmentation process, we were not able to formally verify genre identity, and relied on vocal characteristics to infer gender, introducing a potential source of labeling uncertainty.

All datasets used in this study were sourced from publicly available resources and supplemented with carefully annotated samples to improve representation across groups. We are committed to transparency and reproducibility in our research practices and publish the supplemented metadata alongside this work.

While our focus is on diagnosing and mitigating biases, we also acknowledge the broader ethical implications of our work. This includes the potential misuse of debiasing techniques and the unintended consequences of highlighting biases. Engaging with communities affected by these biases is crucial for ensuring that our research is grounded in real-world experiences and needs.

Our findings are intended to foster critical reflection on the biases and assumptions embedded in music technologies. We hope this work encourages broader research into how MIR models understand and represent music, and the social and cultural implications that follow.

\bibliographystyle{plain}
\bibliography{ISMIR2025_template}

\end{document}